\documentclass[2018, preprint, 12pt, 1p]{elsarticle}

\usepackage{amssymb}

\usepackage{graphicx}
\graphicspath{
	{images/}
}

\usepackage{upgreek}

\usepackage[hyphens]{url}

\journal{arXiv.org}

\begin{document}

\begin{frontmatter}
\title{Rethinking Image Sensor Noise for Forensic Advantage\tnoteref{copyright}}

\author[eee]{R.~Matthews\corref{cor1}}
\ead{richard.matthews@adelaide.edu.au}
\author[eee]{M.~Sorell}
\author[cs]{N.~Falkner}
\cortext[cor1]{Corresponding author}
\tnotetext[copyright]{ Copyright 2018. This manuscript version is made available under the CC-BY-NC-ND 4.0 license http://creativecommons.org/licenses/by-nc-nd/4.0/}

\address[eee]{The University of Adelaide, School of Electrical and Electronic Engineering, Adelaide, SA, 5005 AUS.}
\address[cs]{The University of Adelaide, School of Computer Science, Adelaide, SA, 5005 AUS.}

\begin{abstract}
Sensor pattern noise has been found to be a reliable tool for providing information relating to the provenance of an image. Conventionally sensor pattern noise is modelled as a mutual interaction of pixel non-uniformity noise and dark current.  By using a wavelet denoising filter it is possible to isolate a unique signal within a sensor caused by the way the silicon reacts non-uniformly to light. This signal is often referred to as a fingerprint. To obtain the estimate of this photo response non-uniformity multiple sample images are averaged and filtered to derive a noise residue. This process and model, while useful at providing insight into an images provenance, fails to take into account additional sources of noise that are obtained during this process. These other sources of noise include digital processing artefacts collectively known as camera noise, image compression artefacts, lens artefacts, and image content. By analysing the diversity of sources of noise remaining within the noise residue, we show that further insight is possible within a unified sensor pattern noise concept which opens the field to approaches for obtaining fingerprints utilising fewer resources with comparable performance to existing methods.
\end{abstract}

\begin{keyword}
Sensor Pattern Noise \sep Photo Response Non-Uniformity \sep Digital Image Forensics \sep Dark Current


\end{keyword}

\end{frontmatter}


\section{Introduction} 
Sensor pattern noise (SPN) is a reliable tool for tracking the provenance of images \cite{lukas2006digital}. Through the use of high-pass filtering, a unique signal can be extracted from an image consisting of Photo-Response non-uniformity (PRNU) noise. This signal is unique to the image sensor and is capable of discrimination across cameras of the same make and model.  This discrimination is because the PRNU is defined as the pixel to pixel variance in output intensity of an image sensor when illuminated with a constant light source. The PRNU is the light-sensitive signal caused by Pixel Non-Uniformity (PNU) within a discrete image sensor. It is statistically unlikely for two image sensors to have the same PRNU fingerprint. This capability has been demonstrated with a false acceptance rate of 0.0024\% and a false rejection rate of 2.4\% making PRNU comparison an attractive tool where other evidence can also be used to verify the outcome \cite{goljan2009large}. The PRNU approach was further reinforced experimentally by \cite{farid2016photo} showing that a camera with a positive match to an image with the same PRNU is 1/100,000 or 99.999\%.

Since images in the real world are often never illuminated with a constant light source, solving the blind source camera identification problem in this manner requires large sample sizes of images specially crafted to ensure scene contamination is minimised. Processing the large sample of images is either time consuming or requires large amounts of computing resources to ensure efficacy. These resources are not always available for forensic investigators in the field who need efficient tools to quickly and accurately quarantine evidence. There are further challenges in maintaining chain of evidence for embedded cameras, such as in contemporary smart phones and the emerging field of wearable technologies.

Through careful analysis of the current sensor technology in terms of optical effects, semiconductor physics and the environment image sensors operate in, this paper considers the current methodology for measuring the unique PRNU signal and shows that other options exist for extracting the unique signal that may provide more accurate results and lead to the development of more efficient tools.

\section{Basic operation of an Image Sensor} \label{sec:basicop}

The fundamental principle of collecting a digital image has not changed since the first experiments involving selenium-coated metal plates \cite{SWINTON_1926}. Since then the progression from plates to tubes \cite{1686455} (realising the vision of \cite{SWINTON_1908}), to Charged Couple Device (CCD) arrays \cite{boyle1974buried} \cite{boyle1974three} and currently to CMOS imaging sensors (CIS) \cite{matsumoto1985new} \cite{fossum1993active} has seen the quality of the image improve, but the principle remains the same.  Photons are converted to electrons in a PN junction of a photosensitive material through the recombination of holes and electrons via the photoelectric effect.  In the current state of the art pinned photodiode (PPD) conversion of photons to electrons is done in the heavily P doped layer of the PN diode. This is because the PPD architecture results in the depletion layer being almost the entire width of the P+ region \cite{theuwissen2008cmos}. 

\begin{figure}[!t] 
	\centering
	\includegraphics[width=21pc]{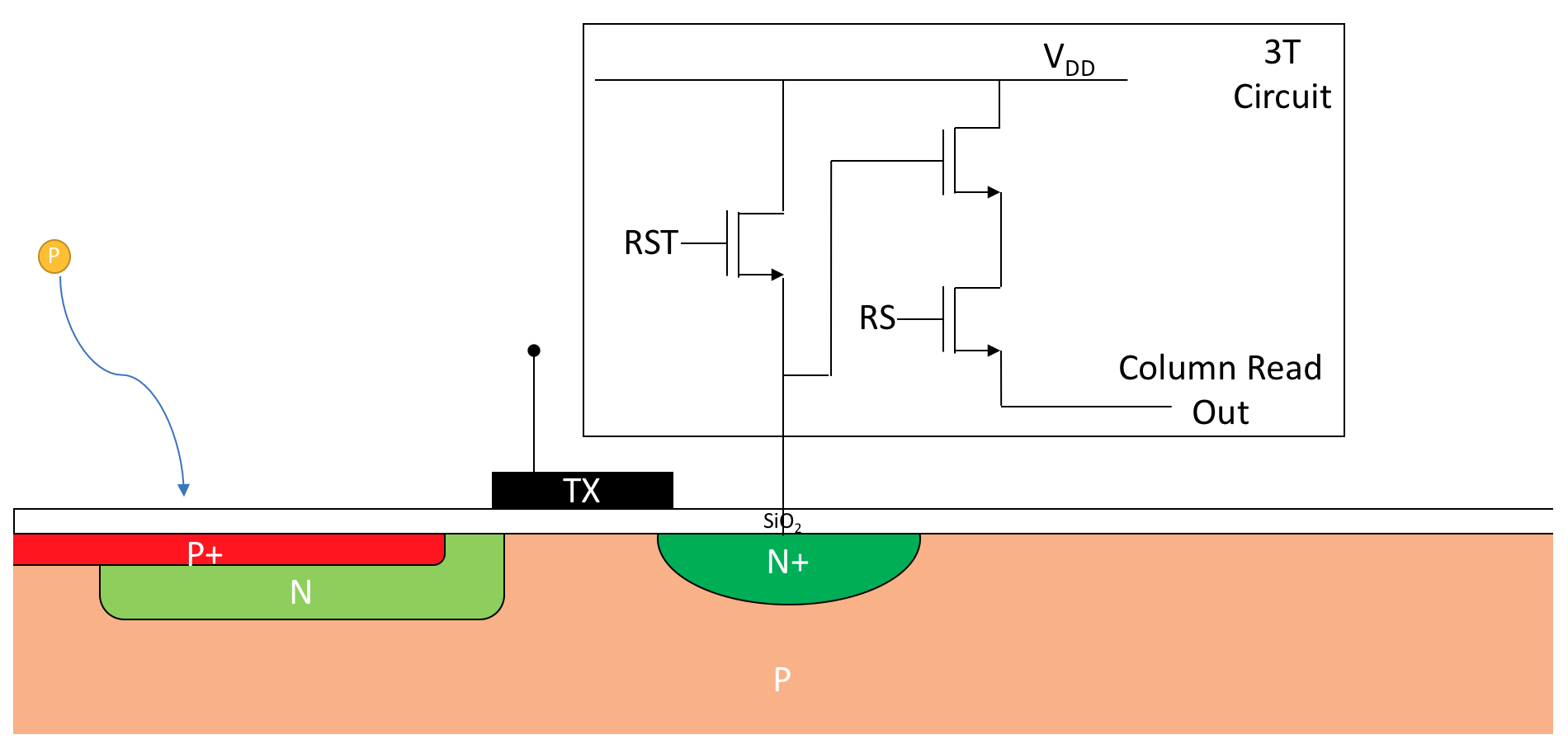}
	\caption[The structure of a Pinned Photo Diode CMOS Image Sensor]{PPD CIS with a three transistor circuit built into the pixel. Each transistor is for a specific function. RST resets the PPD back to full positive voltage at the end of the read cycle to decrease readout noise. RS is used to select the correct pixel in combination with the column bus. TX is used to transfer the charge from the photodiode to the readout node. When a photon P strikes the heavily P doped region P+, the photoelectric effect causes the voltage to decrease across the PN junction. }
	\label{PPD}
\end{figure}

The PPD is the preferred architecture for modern CIS (Figure \ref{PPD}) due to several significant advantages over its predecessors \cite{holst2007cmos}. The PPD exhibits lower noise, lower dark current, higher sensitivity and broader dynamic range than traditional photodiodes or CCDs. As CMOS technology advances we have seen the image sensor shrink in size. However, due to the limitations of CMOS architecture pixel size has been unable to effectively make pixels smaller than 3 $\upmu$m without sharing the on-pixel readout circuity between multiple pixels. This shared pixel concept has allowed pixels to approach the practical limit of 1.0 $\upmu$m \cite{holst2007cmos}. This technology dominates the current generation of mobile phone image sensors and results in a biased increase of fixed pattern noise (FPN) corresponding to the macroblocks of pixels sharing the same transistors.

\begin{figure}[!t] 
	\centering
	\includegraphics[width=21pc]{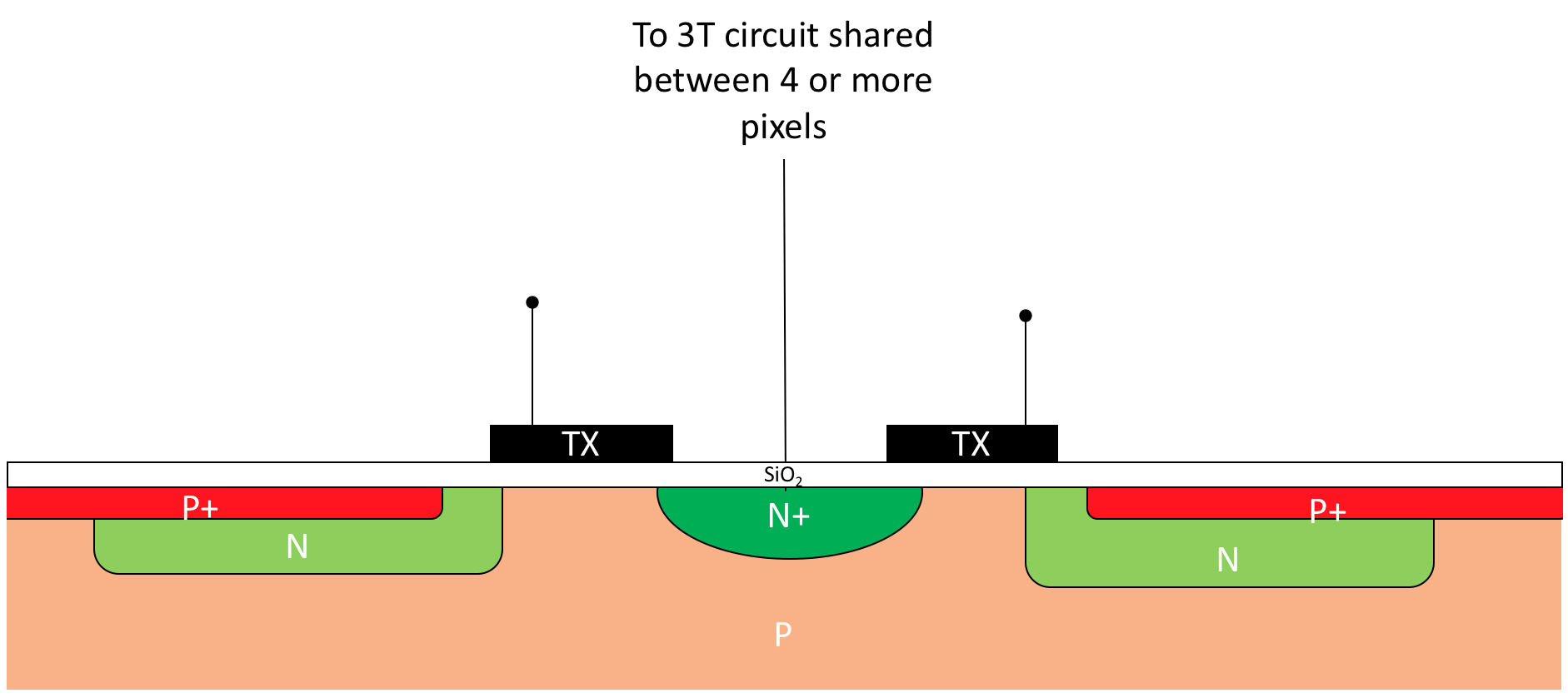}
	\caption[Shared Pixel Concept]{The Shared Pixel Concept results in four or more pixels sharing common readout circuitry allowing pixel pitch to approach the 1$\mu m$ limit. It is theorised that sharing readout circuitry increases Fixed Pattern Noise in macroblocks of a size matching the number of pixels sharing the same circuitry. }
	\label{PPDShared}
\end{figure}

A principle driving factor behind the inability to shrink the photo-sensor area is due to the limitations imposed by the signal-to-noise ratio (SNR) \cite{holst2007cmos}. If consideration is isolated to photon shot noise, the statistical variation of photons striking the sensor, it can be shown that the absorption of incoming photons by a pixel is easily modelled as a Poisson process \cite{holst2007cmos}.   These photons are also characterised by a noise component $\sigma_{ph}$ which is known as shot noise:

\begin{equation}
\sigma_{ph} = \sqrt{\mu_{ph}}
\end{equation}

The flux of $\mu_{ph}$ photons results in $\mu_e$ electrons stored in this pixel since the photoelectric effect causes direct integration of photos to electrons also characterised by a noise component $\sigma_e$, which has a square root relationship with $\mu_e$. 

Assuming a hypothetical noise-free image sensor and noise-free electronics, the performance of the image sensor based system will be limited by photon shot noise. The maximum signal-to-noise ratio can be described as follows:

\begin{equation}
\left( \frac{S}{N} \right) _{MAX} = \frac{\mu_e}{\sigma_e} = \frac{\mu_e}{sqrt{\mu_e}} = \sqrt{\mu_e}
\end{equation}

From this equation, it can be seen that the CMOS process does not determine the minimum size of the pixel. Instead, the number of electrons that can be stored in the pixel successfully while overcoming any noise issues is the determining factor \cite{theuwissen2008cmos}.

Even with these noise limitations however, efforts to shrink pixels continue unabated with new technologies combining CCD and CIS techniques with integrated pixel optics to achieve "subapertures" as small as 0.75 $\upmu$m in size \cite{fife20083mpixel}. 

Accurate noise models, analysis of image sensors regarding noise, and attempts to reduce noise, are thus important areas of research to maximise the efficiency of ever-shrinking image sensors.

\section{Rethinking the noise model for Forensic Advantage}

Currently, the pixel size is not determined by the limitations of CMOS technology but rather the physical capacity for electrons within the N-well region of the photo-detector, be it a photo-gate, photo-diode or pinned photo-diode \cite{theuwissen2008cmos}. This limitation is seen with current CIS  unable to breach the 1.0 $\upmu$m pixel pitch limit \cite{ephotozine} even with CMOS fabrication currently pushing beyond the 1nm scale. The number of electrons that can fit in the pixel is an important observation that will be revisited.

\begin{figure}[!t] 
	\centering
	\includegraphics[width=21pc]{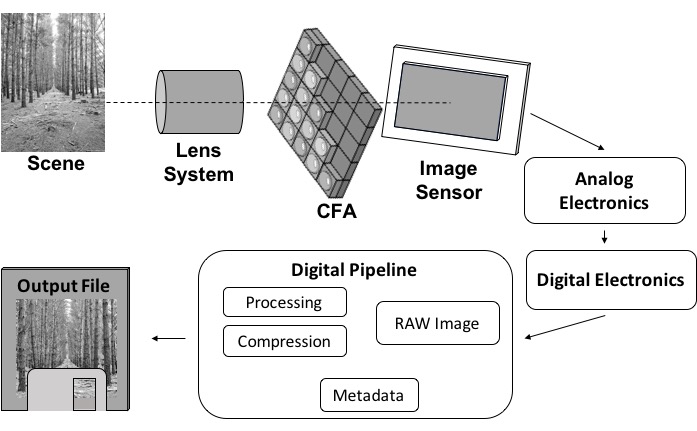}
	\caption{The image capture process referred to as the image pipeline contains two discrete sections, the analogue and the digital. Each element within the pipeline can be exploited to solve the camera identity problem. The analogue pipe consists of the lens optical system including the colour filter array and microlens system, the image sensor and the analogue readout electronics. The digital pipe commences from the output of the analogue to digital converters  (ADC) and involves all the in-camera digital processes the discrete digital signal transverses before finally being saved as an image file. }
	\label{pipeline}
\end{figure}

\begin{figure}[!t] 
	\centering
	\includegraphics[width=21pc]{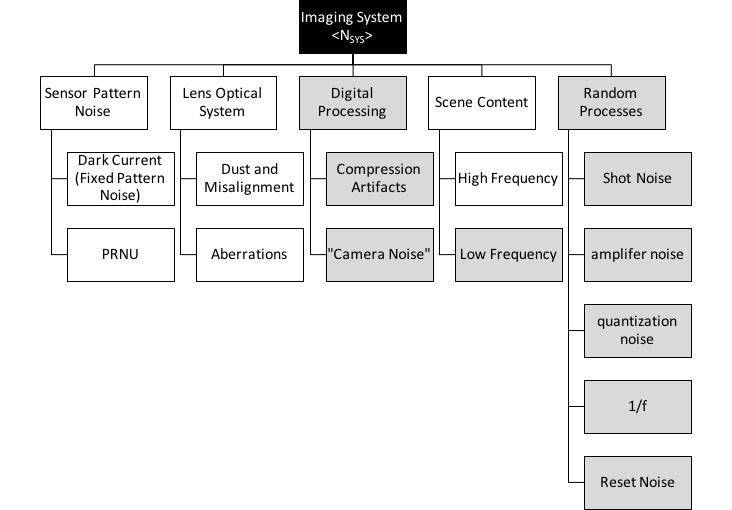}
	\caption[New Additive Noise Model]{The noise residue model as proposed in our work based on the system noise equations from \protect{\cite{holst2007cmos}}. The dark grey boxes indicate sources of noise that can be easily removed. Random processes are traditionally removed through frame averaging \protect{\cite{holst2007cmos}} while RAW format images remove digital processing artefacts\protect{\cite{knight2009analysis}}. The low-frequency components of the scene content and all other sources of noise are removed due to the high-pass filter that the images are passed through to obtain the noise residue in the current unique PRNU signal fingerprinting method .}
	\label{newModel}
\end{figure}

An abstract model of noise within image sensors can be developed by first focusing on what is known as the image pipeline (Figure \ref{pipeline}). This pipeline is the process through which an optical image is converted and processed as an electronic signal to result in a digital image capable of being saved in the multiple formats commonplace today. Each element of the pipe adds an element of noise to the signal resulting in an additive noise model \cite{holst2007cmos}. Through careful analysis of this pipeline, a noise model has been developed (Figure \ref{newModel}). 

For a single image, the discrete sources of noise can be modelled as an additive combination of sensor pattern noise, lens optical effects, digital processing artefacts, scene content and random process \cite{holst2007cmos}. To analyse the noise model for forensic advantage, the focus is linked to the areas that are related to the image sensor itself, namely the optical effects caused by integrated filters and lenses, noises caused by semiconductor physics of the sensor and integrated ``on-chip'' electronics, and the impact of the environment the sensor operates within. Such a focus allows the image sensor to uniquely describe the camera and not the parallel processes an image runs through before saving.

To determine which noise dominates a complete SNR analysis must be undertaken \cite{holst2007cmos}. An SNR analysis is not the focus of this paper and is left as future work. 

\subsection{Optical Effects}

To isolate the image sensor within the model, the first step is to ensure all other contributions of noise from the image pipeline have been eliminated. It is seen in Figure \ref{pipeline} that before entering the sensor, light first must pass through a lens system.  Lens systems are not without error. Aberrations in multi-lens design and the lens itself include spherical aberration, coma, astigmatism, the curvature of field, distortion and chromatic aberration (a particular case of spherical aberration). These are commonly referred to as the primary or third-order Seidel Aberrations after the work of Ludwig von Seidel in 1857 \cite{Seidel_1857}. Each aberration causes the light rays travelling to deviate in some manner from the optical axis of the lens resulting in optical noise which can be confined to either a pixel, group of pixels or the whole image. The relevant mathematical proof behind each aberration has been well explored in the literature, and many texts have been written on the topic \cite{jenkins1937fundamentals}.

\begin{figure}[!t] 
	\centering
	\includegraphics[width=21pc]{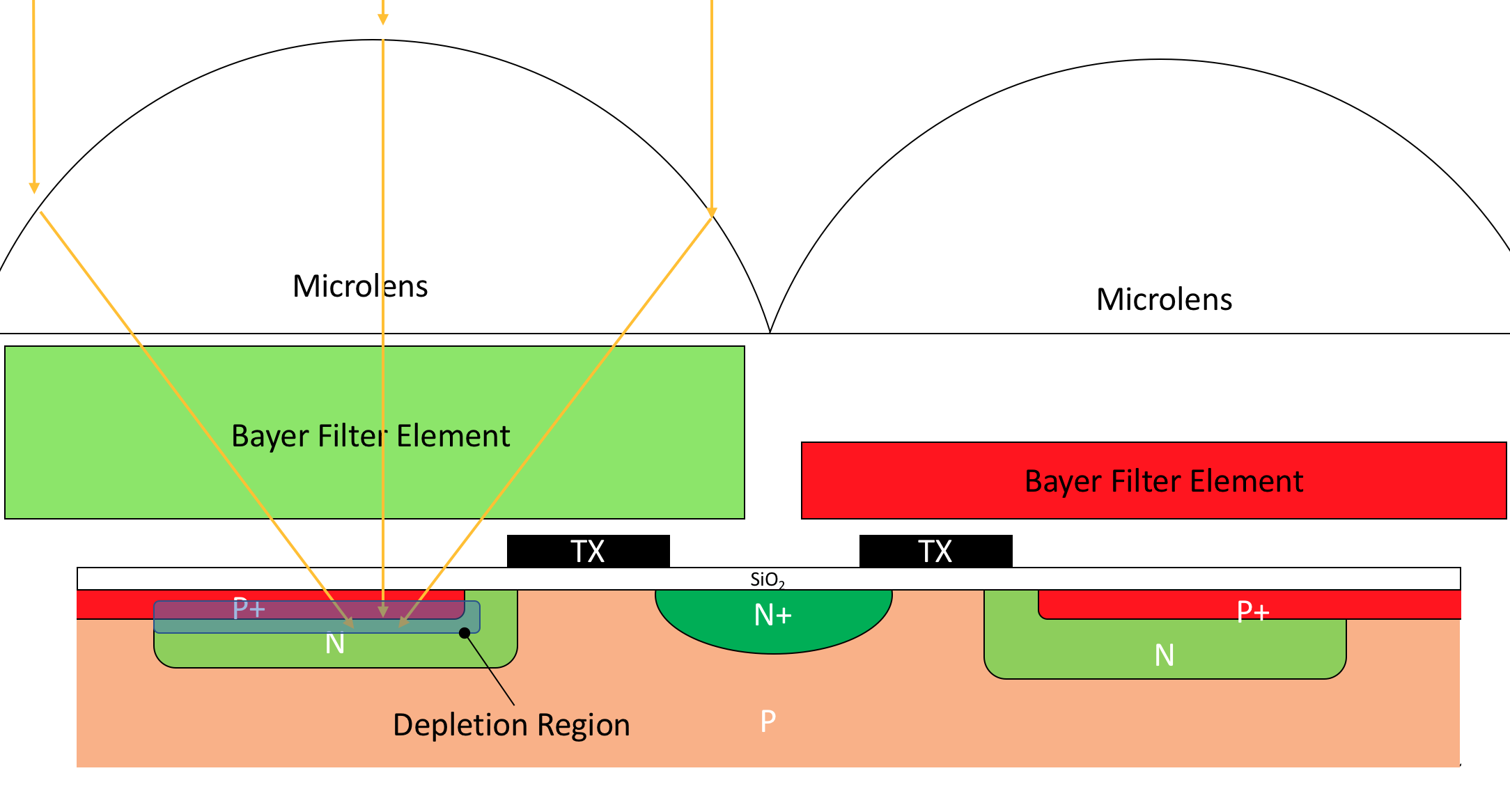}
	\caption[Shared Pixel CIS with CFA and Microlens]{A shared pixel concept CIS with CFA  elements and microlens attached.}
	\label{Optical}
\end{figure}

In addition to the lens system, each image sensor has integrated by design an optical colour filter array (CFA) to ensure colour images can be obtained from broad wavelength light-sensitive silicon. To focus light onto the photosensitive area of the pixel, each pixel also includes a microlens (Figure \ref{Optical}). Should any of these elements be incorrectly manufactured additional noise will be introduced to the system per the same aberrations as above.

Finally, due to physical properties of light, namely Planck-Einstein's Formula, each discrete photon carries different levels of energy according to the wavelength it is travelling at: 

\begin{equation} \label{depth}
E_{ph} = \frac{hc}{\lambda}
\end{equation}

where h is Planck's constant $6.626 \times 10^{(-34)} Js^{-1}$ , c is the speed of light, and $\lambda$ is the wavelength. This energy results in longer wavelengths penetrating deeper into the sensor before being absorbed \cite{holst2007cmos}, affecting photon shot noise.

Optical effects have already been shown to be useful for forensic advantage using discrete lenses and CFA processing artefacts as the identifier. This identification has been achieved primarily through the use of radial lens distortions \cite{san2006source} and CFA interpolation algorithms \cite{bayram2005source}. What has not been shown is how variance in the CFA elements construction may affect penetration of photons to the sensor substrate below. Additionally, aberrations within the integrated microlens have yet to be included in the consideration of the overall noise profile since the sensor contains these elements that physically cannot be removed without destruction. This integrated microlens provides a point of difference that may be exploited for forensic advantage or otherwise contaminate the noise profile of the underlying silicon.

\begin{figure}[!t] 
	\centering
	\includegraphics[width=21pc]{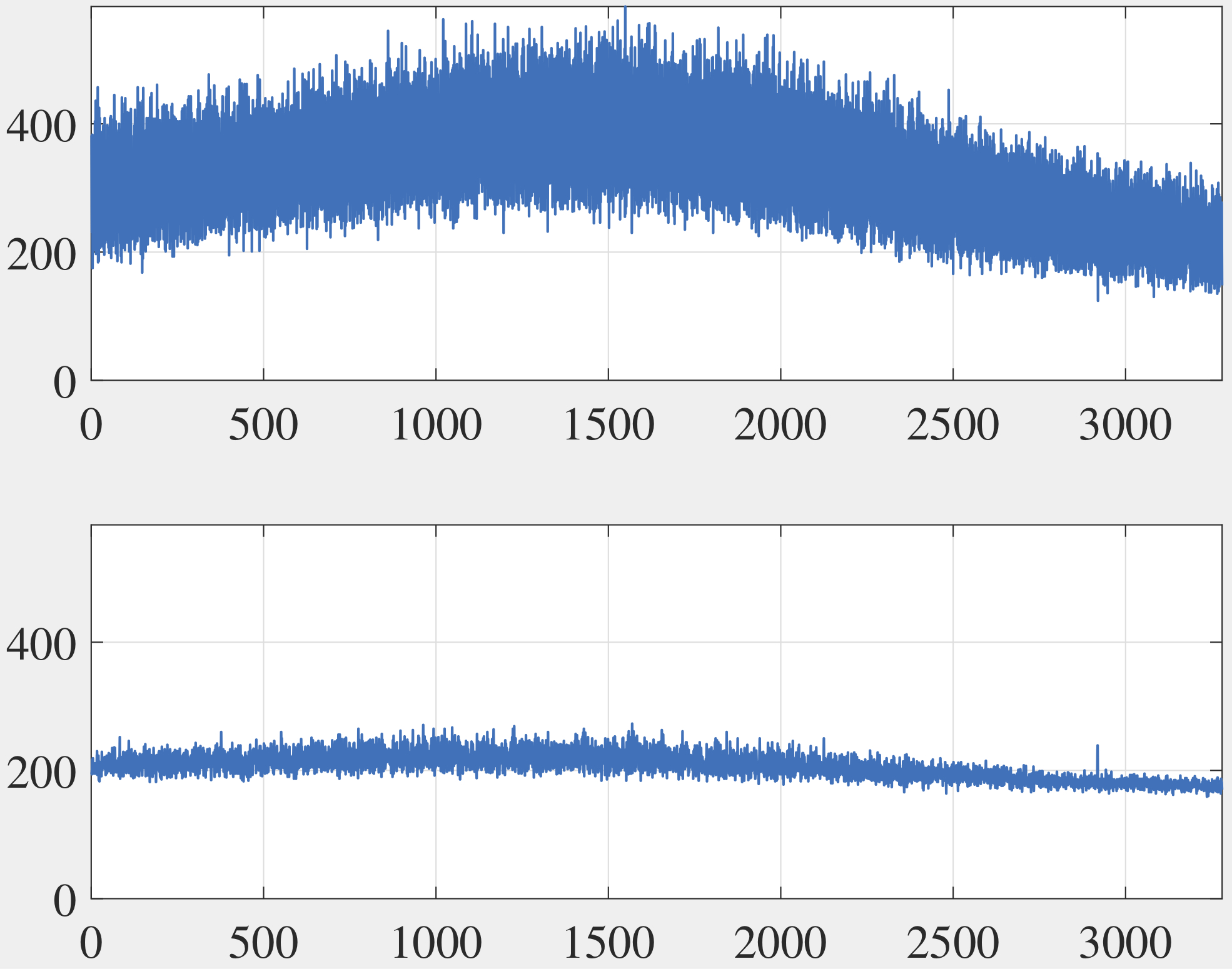}
	\caption{Comparing image line 1024 of a 2048x2048 image we see that an image taken with a lens (top) shows aberration effects with the light sloping towards the centre of the image. Replacing the lens with a pinhole (bottom) removes these aberrations as demonstrated from the decreases in pixel to pixel variance and smooth response across the row.}
	\label{OpticalEvidence}
\end{figure}

To illustrate the optical effects a side by side comparison of pixel intensities across a single row of an image taken with two separate lens systems is shown in Figure \ref{OpticalEvidence}. In this graph, the top figure represents a row 1024 of a 2048x2048 image taken by an integrated lens. The bottom section displays the same row as taken with a pinhole lens. The pinhole image was taken with a suitable exposure time to ensure that the amount of light entering the sensor was the same. It is clear that the optical effects are removed since the variation of the pixel intensities is decreased in the pinhole image. This is most obvious towards the centre of the row where the intensity of the pixel is mostly uniform for the pinhole, however it increases for the lens. Such an aberration would be seen in an image as vignetting. 

\subsection{Semi-Conductor Physics}

The basic operation of a photodiode has been described above in section \ref{sec:basicop}. The primary process is to fill the N-well region of the photodiode with electrons in proportion to the number of photons that have excited the sensor. Under normal operation, however, the number of photon-induced electrons is combined with dark current electrons caused by the physical properties of the PN junction. Three primary sources generate this dark current: irregularities in the silicon structure, diffusion current due to Fick's law and depletion region current which follows Ohm's Law \cite{bogaart2009very}.

The various sources of dark current are complicated to model. This complexity is also in part due to issues with generation in multiple regions as dark current is not just generated in the photosensitive region. These regions include the depletion region, the field-free region and the surface of an oxide layer interface, as well as dark current increasing exponentially with temperature \cite{holst2007cmos}. 

With much work devoted to the subject, it is often enough to model these interactions by the following \cite{holst2007cmos}:

\begin{equation}\label{DarkCurrent}
D_{e^-} = \frac{J_{D}A_{D}t_{int}} {q}
\end{equation}

Where $A_{D}$ is the detector area, $t_{int}$ is the integration or exposure time, $q$ is the charge of an electron $q = 1.6 \times 10^{-19}$ coul and $J_{D}$ is the dark current density which is proportional to:

\begin{equation}\label{DarkCurrentDensity}
J_{D} \propto T^{2}e^{\frac{(E_{t}-E_{G})}{kT}}
\end{equation}

where $k$ is Boltzmann's constant, $k = 1.38 \times 10^{-23}$ J/K, $T$ is the temperature in degrees Kelvin and ($E_{t}-E_{G}$) is the difference in bandgap energies for the impurity carrier and the primary carrier respectively.

From here it is possible to estimate the dark current at a specific temperature from a known image sensor at a given exposure time.

These dark current electrons $n_{DARK}$ will be combined with the photon generated electrons $n_{PE}$to fill the N-well region:

\begin{equation} \label{equa:darkcurrent}
n_{PE} + n_{DARK} = n_{WELL}
\end{equation}

Since no silicon wafer is without defect and no two pixels are uniformly the same it is seen that the dark current will be measurably different between two pixels, however, is usually treated as uniform and quasi-stable for a sensor as a whole \cite{smith2008pixel}.

At the heart of the operation of an image sensor is the N-well region filling with electron generated photons thanks to the photoelectric effect at the depletion region between a PN junction at the N-well region. Since the size of this well differs from pixel to pixel, and sensor to sensor, it is possible to create a unique signal or fingerprint from the PRNU noise. This concept is the work of \cite{lukas2006digital} which focuses on how N-well of each pixel can be filled to an equal amount via photonic energy. Each pixel is then read out via the well described processes and ultimately saved as an image. The slight variation of each pixel is measured on a pixel to pixel basis due to the differences in the ability of the PN junction of the photosensitive region to recombine photons. 

It has been stated that dark current can be ignored for forensic purposes due to dark-frame removal \cite{lukas2006digital}, the subtraction of a frame exposed without opening the shutter of the same length of time immediately before taking an image. However, since the N-well region can be filled with electrons via dark current generation (equation \ref{equa:darkcurrent}), it is theoretically possible to measure a unique PNU signal in the same manner with the dark current being the excitation source rather than photons. 

\begin{figure*}[!t] 
	\centering
	\includegraphics[width=\textwidth]{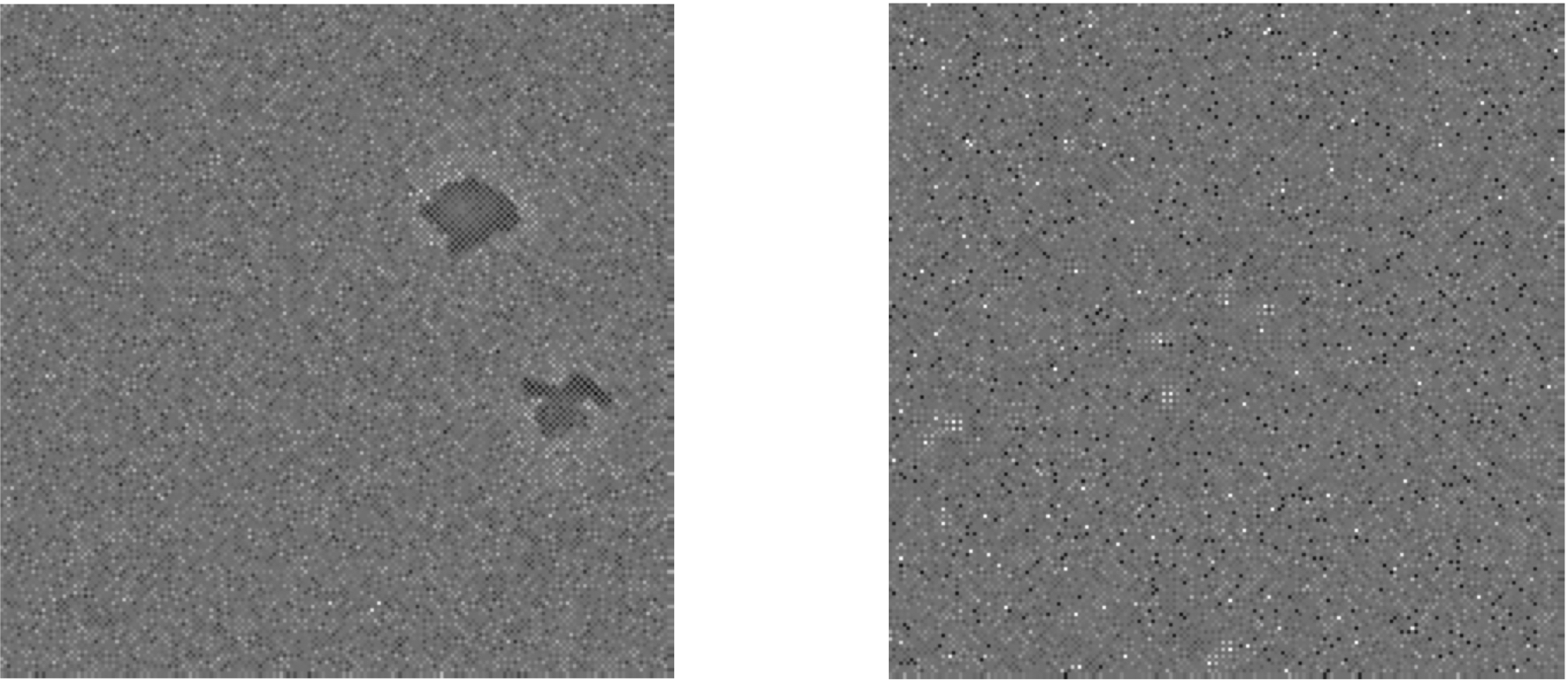}
	\caption{A side by side comparison of the green layer of the fingerprint obtained using PRNU through a pinhole (left) and dark  current (right). The effects from sensor dust are clearly visible in the pinhole image while these are absent in the dark current equivalent section. Kurosawa hot pixels are apparent in the dark current fingerprint as salt and pepper noise on top of the fingerprint.}
	\label{PRNUvDARK}
\end{figure*}

Since dark current density increases exponentially with temperature and proportionally with exposure time it is theoretically possible to completely saturate the N-well region with electrons generated purely from dark current, especially if the pixel pitch is small, as seen in mobile devices. By controlling these two parameters, it is proposed that a valid unique fingerprint can be obtained from a sensor using dark current electrons alone. Such a fingerprint is demonstrated in Figure \ref{PRNUvDARK}.  Demonstrating the reliability of such a fingerprint in a forensic context is beyond the scope of this paper and will be demonstrated in future work. However, to demonstrate that the fingerprints are indeed similar some observations are made. Within the dark current fingerprint the Kurosawa hot pixels \cite{kurosawa1999ccd} are required to be suppressed for an accurate comparison. This results in a salt and pepper noise artefact present within the fingerprint. Additionally, since light is not used to generate the fingerprint there is no contamination from dust as seen in the PRNU reference pattern.
\begin{table*}[!t]
	\renewcommand{\arraystretch}{1}
	\caption {Correlation of 100 image PRNU reference pattern vs single dark current fingerprint: Camera One}
	\centering
	\label{CameraOne}
	\begin{tabular}{ccccc}
		\hline
		Temp ($^{\circ}$C) & 0$^{\circ}$ Rotation  & 90$^{\circ}$ Rotation & 180$^{\circ}$ Rotation & 270$^{\circ}$ Rotation \\
		\hline
		20 & 0.0219 & 1.30e-03 & 2.00e-03 & 1.70e-03  \\
		
		45 & 0.0408 & -8.68e-04 & 1.28e-05 & 3.50-05   \\
		\hline                      
	\end{tabular}
\end{table*}

\begin{table*}[!t]
	\renewcommand{\arraystretch}{1}
	\caption{Correlation of 100 image PRNU reference pattern vs single dark current fingerprint: Camera Two}
	\centering
	\label{CameraSix}
	\begin{tabular}{ccccc}
		\hline
		Temp ($^{\circ}$C) & 0$^{\circ}$ Rotation  & 90$^{\circ}$ Rotation & 180$^{\circ}$ Rotation & 270$^{\circ}$ Rotation \\
		\hline
		20 & 0.0335 & 1.10e-03 & 2.00e-03  & 7.23e-05 \\
		
		45 & 0.0740  & -7.56e-04   & 1.28e-04   & -2.64-08    \\
		\hline                      
	\end{tabular}
\end{table*}

\begin{table*}[!t]
	\renewcommand{\arraystretch}{1}
	\caption{Correlation of 100 image PRNU reference pattern vs single dark current fingerprint: Flipped PRNU}
	\centering
	\label{Flipped}
	\begin{tabular}{ccccc}
		\hline
		Temp ($^{\circ}$C) & 0$^{\circ}$ Rotation  & 90$^{\circ}$ Rotation & 180$^{\circ}$ Rotation & 270$^{\circ}$ Rotation \\
		\hline
		20    & 1.30e-03 & -8.01e-04 & 7.67e-04 & 1.30e-03 \\
		
		45 & 8.733-04 & -1.20e-03 & -7.77e-05 & -9.31-04 \\
		\hline                      
	\end{tabular}
\end{table*}

Demonstrating that the dark current fingerprint is correlating to the PRNU fingerprint a set of correlations is calculated for a single image dark current fingerprint at T=20$^{\circ}$C and 45$^{\circ}$C against a 100 image PRNU reference pattern (Table \ref{CameraOne}). The PRNU reference pattern is rotated 90$^{\circ}$, 180$^{\circ}$ and 270$^{\circ}$ as a proxy for a deliberate mismatch to ensure correlation is indeed occuring with the reference pattern and not an arbitary artefact of the sensor design. This is repeated for another camera in table \ref{CameraSix} with the results confirming the previous demonstration. Finally, to ensure that the correlation is between the excitation of the silicon and not the read out of the sensor design itself half the PRNU reference pattern is swapped with the other half of the reference pattern to create a deliberate mismatch while maintaing colour filter rotation. These results are shown in table \ref{Flipped} with no significant correlation detected. These results demonstrate our hypothesis that a dark current fingerprint can be generated which may form as a substitute for the current PRNU methodologoy. A full analysis is left as future work.

\section{Micrometer Imagery}

\begin{figure}[!t] 
	\centering
	\includegraphics[width=21pc]{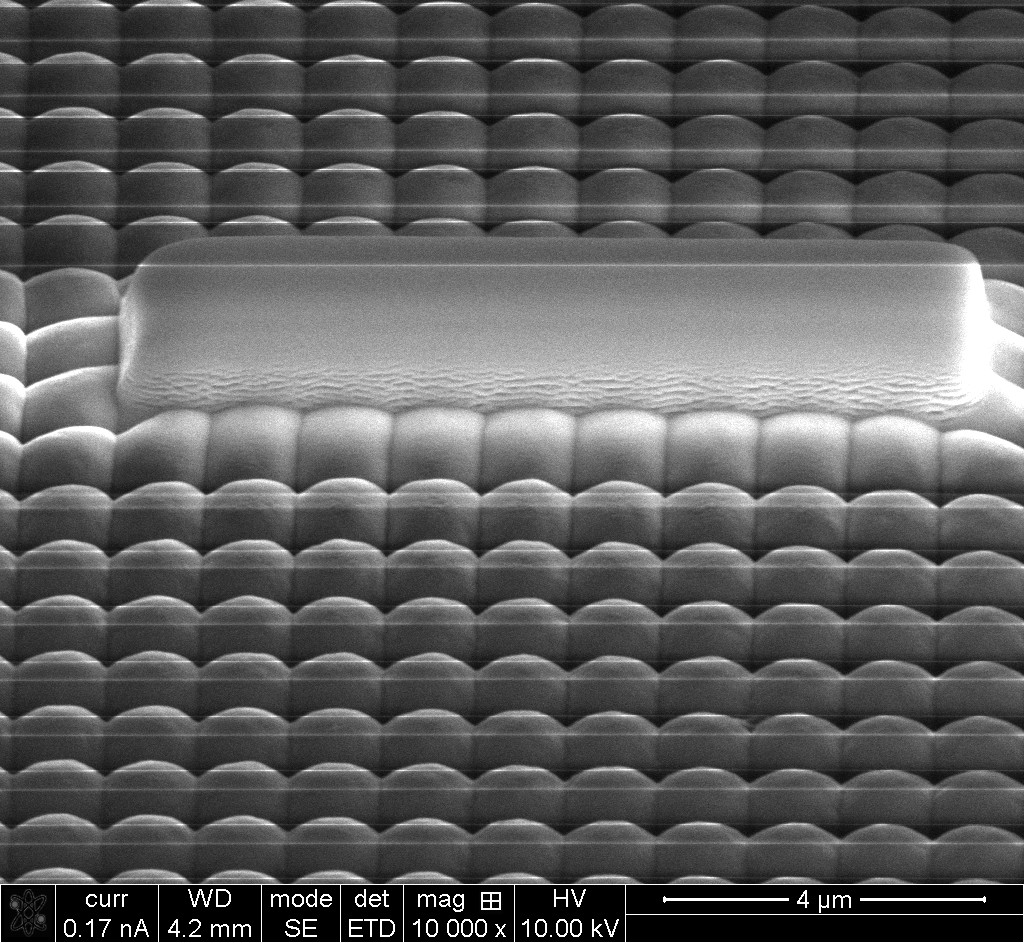}
	\caption[Platinum deposit on CIS]{Platinum (shown here as a growth on top of the micro-array) is deposited on the CIS to prevent micro-fractures forming during the excavation process.}
	\label{Platinum}
\end{figure}

\begin{figure}[!t] 
	\centering
	\includegraphics[width=21pc]{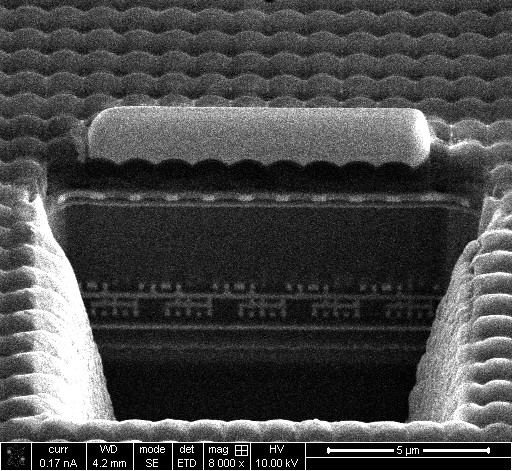}
	\caption[Excavated CIS]{The excavated region of the CIS is shown after the Gallium ion beam has been used to step out the material present in the region of interest. Several passes are used to obtain a smooth, polished cross-section.}
	\label{excavation}
\end{figure}

To illustrate these concepts we have discussed above a cross-sectional image of a Sony IMX219PQ \cite{Sony_2018} image sensor was taken under a scanning electron microscope. Using an FEI Dual Beam Helios Nano Lab 600 \cite{fei} scanning electron microscope, two excavations were made into a Sony IMX219PQ CIS. Since the CIS is conductive, no sample preparation is required before scanning. First, a layer of platinum is deposited above the area to be excavated to prevent fracturing (Figure \ref{Platinum}).  After the platinum is deposited a process of staged cuts are made using a gallium ion beam to create an excavated area through the sensor that can be imaged (Figure \ref{excavation}).

Using a magnification of 20,000x, a current of 0.17nA and voltage of 10kV, images are then obtained of the top layers of the integrated CIS containing the pinned photodiode. Since our study is not concerned with the readout circuitry, we exclude it from our observations. The process is repeated for a diagonal cross section to ensure multiple pixels across CFA regions are obtained.

\begin{figure*}[!t] 
	\centering
	\includegraphics[width=\textwidth]{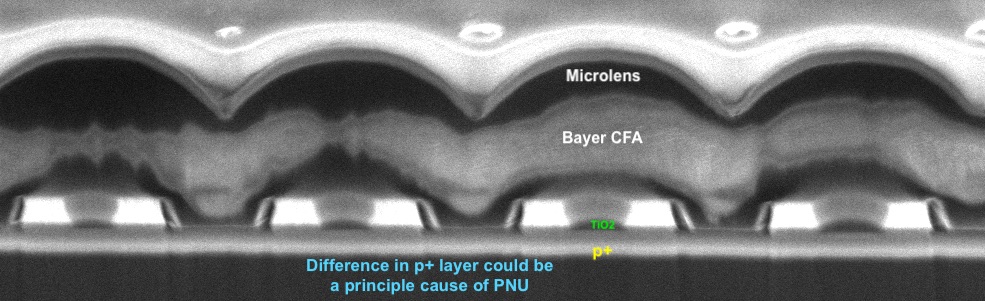}
	\caption[Micrometer imagery of Sony IMX219PQ]{A diagonal cross-sectional view of the Sony IMX219PQ CIS. Four pixels are shown. The pinned photo-diode shared pixel architecture is visible. The Pinning layer is marked P+.}
	\label{Micro}
\end{figure*}

In Figure \ref{Micro} a layered PPD architecture is seen as expected. This PPD architecture uses a shared pixel concept with multiple transfer points (shown as TX). Using Energy-dispersive x-ray spectroscopy, the architecture can be determined in detail. Elements detected in the EDX analysis include Platinum, Carbon, Gallium, Oxygen, Silicon, Tungsten, and Titanium. The presence of gallium and platinum must be excluded since they are used in the  SEM processes outlined above. However, using reasonable assumptions, the areas of the CIS in the image as shown can be reverse engineered. An isolation oxide layer directly below the PN junction made from titanium dioxide is observed as part of the structure. This TiO2 layer electrically isolates the PN junction from the underlying substrate and also isolates any photons from further penetrating into the underlying substrate \cite{holst2007cmos}.  

To explain the structure and operation we refer to \cite{Sony_2018}. Light first enters the sensor via the microlens array and is then filtered using a Bayer filter. The IMX219PQ sensor has a traditional ''R, G, and B primary colour pigment mosaic filters'' CFA \cite{Sony_2018}. These filter elements are not uniform in their construction, and it is seen that even cells of the same colour have different widths (Figure \ref{Micro}). From Equation \ref{depth} it is clear that this will result in different wavelengths being filtered out on a pixel by pixel basis rather than just the three principal components being isolated by the chemical composition of these layers. This filtering will cause chromatic distortion.  While the cause of variation in the CFA layer is impossible to determine from this image, an indication is given from the P+ pinning layer directly below. Since the CMOS process manufactures each layer on top of the previous layer, variations in the layers below will cause issues in the layers above. It is seen that the width of the pinning layer is different from pixel to pixel in the cross-sectional view not just in width but also in length. This pinning layer will affect the performance of the photosensitive PN region and even the dark current of each pixel. 

Since there are variations between each pixel, it may be possible to to isolate and hence exploit these variations, to create better processes to isolate a unique fingerprint for the sensor.

\section{Conclusion}

SPN has shown to be a promising area of research for answering provenance questions relating to imagery. It still suffers from reliance on large data sets ideally constructed from flat fielded images and a-priori information that is not always apparent or readily available to forensic investigators in the field. By rethinking the noise model of a digital camera, SPN can be isolated as an element that is primarily dependent on the physical silicon that each image sensor is built from, regardless of technology. From this analysis, it can be seen that there are alternative ways to create the unique fingerprint that will not rely on large data sets or mass computational resources. An example, the subject of current work, is the exploitation of dark current.

\section{Acknowledgements}
This research did not receive any specific grant from funding agencies in the public, commercial, or not-for-profit sectors. The authors acknowledge the facilities, and the scientific and technical assistance, of the Australian Microscopy and Microanalysis Research Facility at Adelaide Microscopy, the University of Adelaide.
This research is supported by an Australian Government Research Training Program (RTP) Scholarship and forms part of a thesis chapter.




  \bibliographystyle{elsarticle-num} 
  \bibliography{arXivRethinking}

\end{document}